\def\wn{\omega_n}
\def\wj{\omega_j}

\documentclass{elsart}

\usepackage{amssymb}
\usepackage{graphicx}

\begin{document}

\begin{frontmatter}

\title{The limits of the rotating wave approximation in the electromagnetic
field propagation in a cavity}

\author{I. Dolce,}
\author{R. Passante}
\ead{roberto.passante@fisica.unipa.it}
\author{and F. Persico}

\address{CNISM and Dipartimento di Scienze Fisiche ed Astronomiche,
Universit\`{a} degli Studi di Palermo, Via Archirafi 36, I-90123 Palermo, Italy}

\begin{abstract}
We consider three two-level atoms inside a one-dimensional cavity,
interacting with the electromagnetic field in the rotating wave
approximation (RWA), commonly used in the atom-radiation interaction.
One of the three atoms is initially excited, and the
other two are in their ground state. We numerically calculate
the propagation of the field spontaneously
emitted by the excited atom and scattered by the second atom, as well as the
excitation probability of the second and third atom.
The results obtained are analyzed from the point of view of relativistic
causality in the atom-field interaction.
We show that, when the RWA is used, relativistic causality
is obtained only if the integrations over the field frequencies are extended to $-\infty$; on the
contrary, noncausal tails remain even if the number of field modes is increased.
This clearly shows the limit of the RWA in dealing with subtle problems such as
relativistic causality in the atom-field interaction.
\end{abstract}

\end{frontmatter}

\section{\label{sec:1}Introduction}

The propagation of electromagnetic signals has been a subject of
investigation since the beginning of the quantum theory of the
electromagnetic field. In the last years this subject has received much attention
in the framework of rigorous proofs of relativistic causality in
the atom-field interaction \cite{PT83,Berman04,MJF95,CPPP95},
of the recently observed slow light propagation \cite{HHDB99}
and of the possibility of superluminal propagation of light \cite{Chiao93}.

Recently, the propagation and the scattering of a photon
spontaneously emitted by an atom in a one-dimensional cavity has
been investigated using the rotating wave approximation (RWA)
\cite{PTL03}. This approximation consists in neglecting non-energy
conserving terms in the Hamiltoniam, and the potential dangers of
this approximation in terms of the causal behaviour of the
physical system considered are well known \cite{BCPPP90,MJF95}.
Quite frequently, the effect of the RWA seems to be {\it
compensated} by the extension to $-\infty$ of the integrations in
the frequency of the field modes, as in the original Fermi model
description of causality in the excitation transfer between two
atoms, one of which initially excited \cite{Fermi32,UD02,HU04}.
Yet, this is in any case conceptually unsatisfactory for a subtle
problem such as relativistic causality, and much effort has been
dedicated to the inclusion of the counterrotating terms in order
to have a rigorous proof of relativistic causality in quantum
electrodynamics \cite{CPPP95,PT95}.

In this letter we consider the same system considered by Purdy
{\it et al.} in \cite{PTL03}. This system consists of three
two-level atoms inside a one-dimensional cavity, interacting with
the electromagnetic field in the RWA. We show numerically that the
causal behaviour obtained in \cite{PTL03} (using the RWA) indeed
derives from an extension to $-\infty$ of the frequency
integrations; this extension, however, has no physical
justification since it makes the Hamiltonian unbounded from below,
and this is in general a fundamental point for the causality
problem, as pointed out by Hegerfeldt \cite{Hegerfeldt94}. We show
that even if the Hamiltonian is bounded from below, noncausal
terms in the field propagation are present when the rotating wave
approximation is used; these terms do not vanish when the number
of field modes is increased. This explicitly shows that the RWA is
not an appropriate approximation for dealing with problems of
relativistic causality in matter-radiation systems, and that the
counterrotating terms should be included.

Our system is the same as in \cite{PTL03}: it consists of three
two-level atoms, named 1, 2, 3, in a one-dimensional cavity. The
cavity has length $L$, with two parallel plates at $x=0$ and
$x=L$, and the positions of the atoms 1, 2, 3 are $L/4$, $L/2$ and
$3L/4$, respectively. In the Coulomb gauge and multipolar coupling
scheme, within dipole approximation and using the rotating wave
approximation, the interaction of the three atoms with the
radiation field is described by the following Hamiltonian (with
units such that $\hbar = 1$)
\begin{equation}
H = \sum_{j=1}^3 \wj S_j^z + \sum_n \wn a_n^\dagger a_n +
\sum_n \sum_{j=1}^3 \left( g_{jn} a_n S_j^+ + g_{jn}^\star a_n^\dagger S_j^- \right)
\label{eq:1}
\end{equation}
where index $n=1,2,3, \ldots$ denotes the field modes of the
cavity and $j=1,2,3$ denotes the atoms. $a_n, a_n^\dagger$ are the
annihilation and creation operators of the n-th mode, $S_j^\pm ,
S_j^z$ are the pseudospin operators of the atom $j$, and
\begin{equation}
g_{jn} = \Omega_j \sin \frac {n\pi x_h}L \hspace{5pt} ; \hspace{15pt}
\Omega_j = \mu_j \sqrt{\frac {\wj}{2\epsilon_0 L}}
\label{eq:2}
\end{equation}
where $\wj$ is the transition frequency of atom $j$ and $\mu_j$ is
its electric dipole moment. In the expression of the coupling
constant the near-resonance approximation $\wn \simeq \wj$ has
been used.

We wish to stress that, because we are describing our system in
the multipolar coupling scheme, the field operator that we
calculate, the momentum conjugate to the vector potential, is the
transverse displacement field ${\bf D}_\perp ({\bf r})$. This
operator, outside the sources, coincides with the total electric
field ${\bf E}({\bf r})$. This is an essential point, because
${\bf E}({\bf r})$ satisfies a retarded wave equation, and thus it
is expected to manifest causal propagation in space. On the
contrary, when the minimal coupling scheme is used, the field
operator conjugate to the vector potential is the transverse
electric field ${\bf E}_\perp ({\bf r})$, which is not a retarded
operator; in fact, the source term in the corresponding Maxwell
equation is the transverse current density, which is not localized
in space for atomic systems \cite{CPP95}.

Our initial state is the state with atom 1 excited, atoms 2 and 3 in their ground states
and the field in the vacuum state, that is $\mid e, g, g, 0_k \rangle$.
In the RWA the only states participating to the evolution of the initial state are, with an
obvious meaning of the symbols,
$\mid g, e, g, 0_n \rangle$, $\mid g, g, e, 0_n \rangle$,
$\mid g, g, g, 1_n \rangle$.
Thus we can write the general state at time $t$ as the following superposition
\begin{eqnarray}
\mid \psi (t) \rangle &=& c_1(t) \mid e, g, g, 0_n \rangle + c_2(t) \mid g, e, g, 0_n \rangle
\nonumber \\
&+& c_3(t) \mid g, g, e, 0_n \rangle + \sum_n b_n(t) \mid g, g, g, 1_n \rangle
\label{eq:3}
\end{eqnarray}

The Schr\"{o}dinger equation gives the following set of coupled differential equations
for the coefficients

\begin{eqnarray}
\dot{c}_j(t) &=& -i \left( \wj c_j(t) + \sum_n g_{jn} b_n(t) \right)
\label{eq:4a} \\
\dot{b}_n(t) &=& -i \left( \wn b_n(t) +\sum_j^3 g_{jn}^\star c_j(t) \right)
\label{eq:4b}
\end{eqnarray}

We assume that the atoms 1 and 3 have the same transition
frequency, $\omega_1 = \omega_3$, and we indicate with $\delta =
\omega_1 - \omega_2$ the detuning of atom 2 compared with atoms 1
and 3. We also put $L/c = 1$, which means using as the unit time
the time taken by the light to cross the cavity.

Approximate analytical solutions of these equations have been
obtained \cite{PL03}, using a method based on Laplace transforms
\cite{SG72}. We have obtained numerical solutions of these
equations. We integrate numerically the set of differentially
equations (\ref{eq:4a},\ref{eq:4b}) with the
Adams-Moulton-Bashfort method \cite{PTVF92,Wheatley95}. This
multistep method is an algorithm more sophisticated than the
Runge-Kutta method, typically used for this kind of problems,
allowing to obtain more accurate results and to use a much larger
number of cavity modes. In order to facilitate comparison of our
numerical results with the analytical results obtained in
\cite{PL03}, we shall adopt the same numerical values of the
detuning $\delta$ and of the atomic decay rates $\gamma_j = \mid
\Omega_j \mid^2$. Therefore, we use $\gamma_1 = 1$, $\gamma_2 =
16$, $\gamma_3 = 256$ and $\delta = 4$ (in our units).

We calculate the expectation value of the square of the electric field, which is proportional to
the electric part of the field energy density, given by (zero-point terms have been neglected)

\begin{equation}
\langle E^2(x,t) \rangle = 2\frac {\omega_1}L \left| \sum_n b_n(t) \sin \left(
\frac {\wn x}c \right) \right|^2
\label{eq:5}
\end{equation}

Our first numerical calculation involves  a set of equally spaced
field modes, symmetric with respect to the resonance frequency of
the first atom. This simulates an integration extended from
$-\infty$ to $+\infty$, therefore a field Hamiltonian not bounded
from below (as that used in \cite{PTL03}). In this paper we report
the results of two numerical calculations: the first uses $10^4$
modes, 5000 above and 5000 below the atomic frequency; the second
uses $2 \cdot 10^4$ modes, $10^4$ above the atomic frequency and
$10^4$ below. We then compare the results obtained with those of
an analogous numerical calculation in which the field modes are
not symmetric with respect to the atomic frequency, and only the
upper cut-off is increased with increasing the number of the field
modes; this second case simulates a field Hamiltonian bounded from
below. This comparison allows us to understand the role of the RWA
in the behaviour of our system, in particular from the point of
view of relativistic causality.

\begin{figure}[ht]
\begin{center}
\includegraphics*[width=15cm]{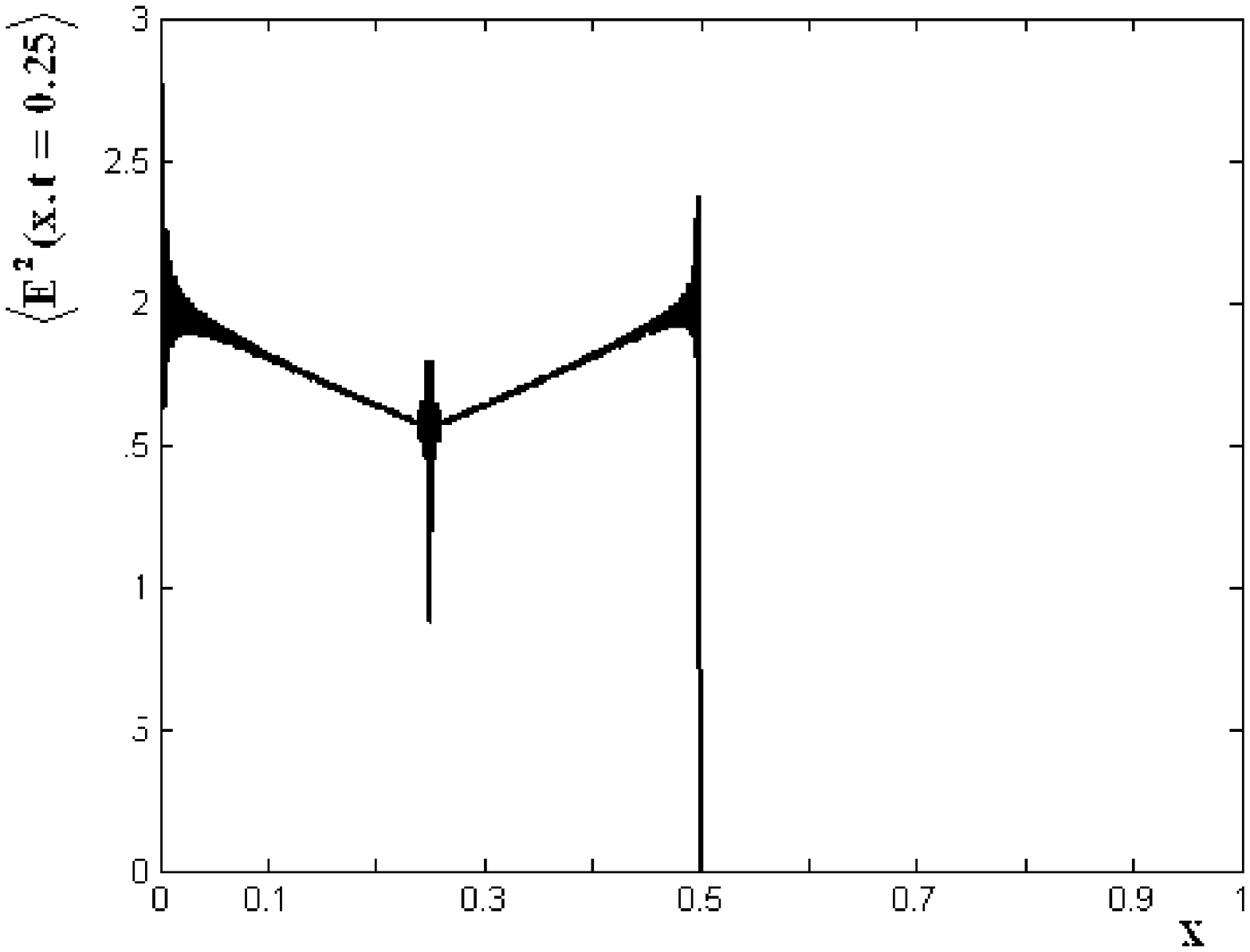}
\end{center}
\caption{Average value of the square of the electric field
$E^2(x,t)$ in arbitrary units at $t=0.25$, with $10^4$ field modes
symmetrically distributed around the atomic transition frequency.}
\label{fig:1}
\end{figure}

\begin{figure}[ht]
\begin{center}
\includegraphics*[width=15cm]{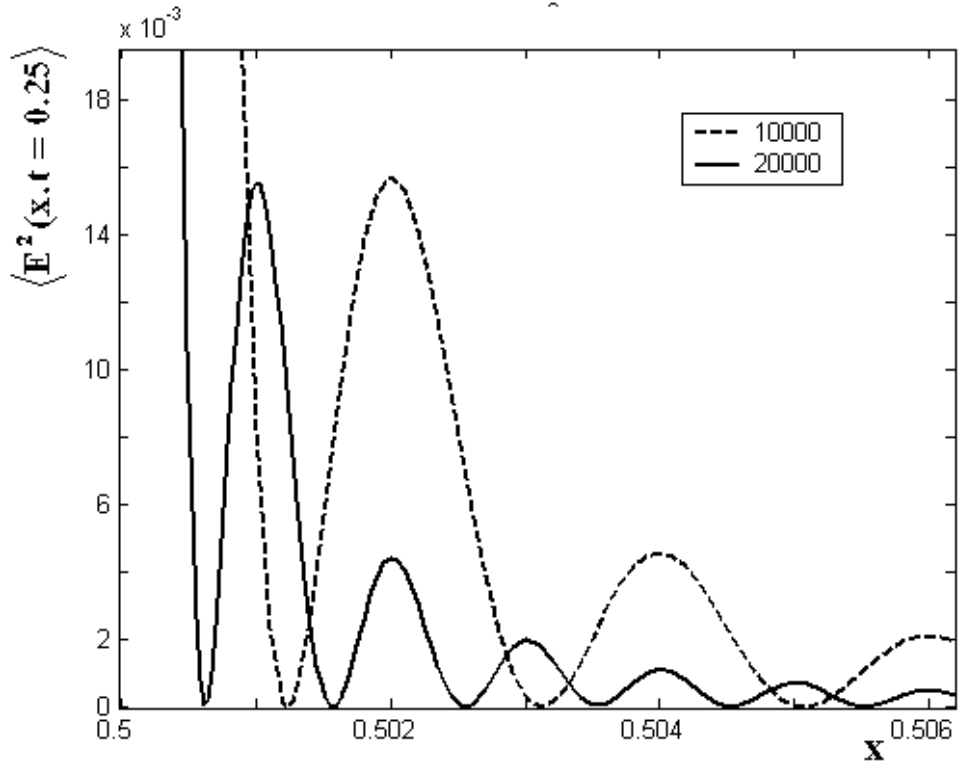}
\end{center}
\caption{Average value of $E^2(x,t)$ in arbitrary units at
$t=0.25$ in the neighbourhood of $x=0.5$. The dashed line is with
$10^4$ modes of the field and the continuous line with $2 \cdot
10^4$ modes (both with a symmetric mode distribution around the
atomic transition frequency).} \label{fig:2}
\end{figure}

Fig. \ref{fig:1} shows $\langle E^2(x,t) \rangle$ at $t=0.25$ when
$10^4$ modes are used with a symmetric distribution around the
transition frequency. A front in the propagation of the energy
density is evident at $x=0.5$, as expected (the atom that emits
the radiation is located at $x=0.25$); however, a zoom of Fig.
\ref{fig:1} in the neighbourhood of $x=0.5$, given in Fig.
\ref{fig:2}, shows the presence of tails for $x>0.5$, blurring the
front. Fig.  \ref{fig:2} also shows the same expansion around
$x=0.5$ when a larger number of modes is used ($2 \cdot 10^4$),
again with a symmetric configuration around the atomic frequency.
An improvement of the behaviour from the point of view of
causality is evident, with a manifest decrease of the tails for
$x>0.5$. We have also obtained similar results for different
times, showing that the front gets sharper when the number of
field modes is increased, if the frequencies of the field modes
are symmetric around the transition frequency of the atom. This
suggests that the expected causal behaviour is indeed approached
in the limit of an infinite number of field modes symmetrically
distributed around $\omega_1$.

\begin{figure}[ht]
\begin{center}
\includegraphics*[width=15cm]{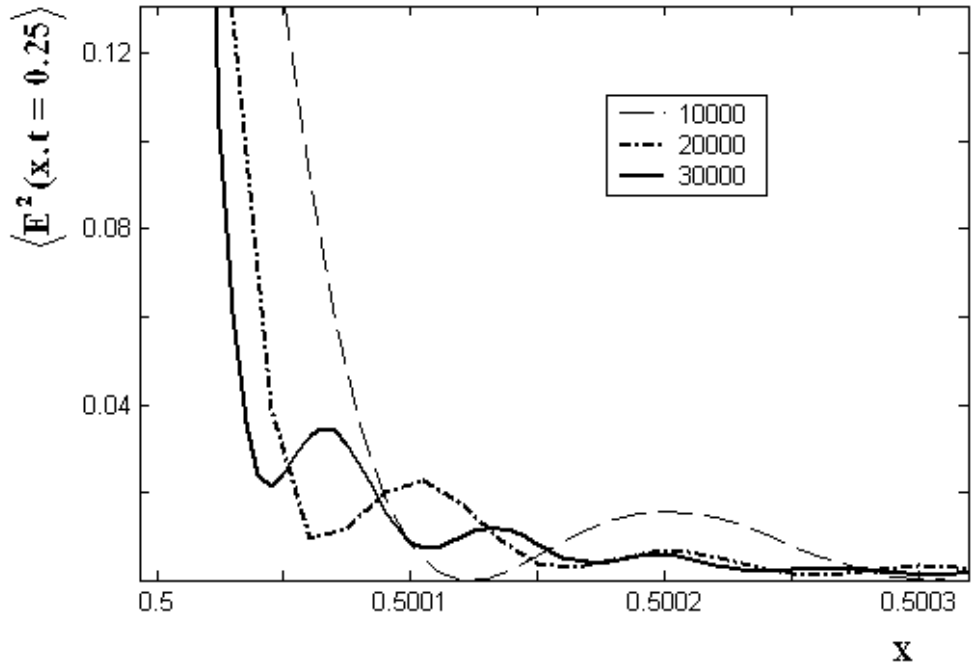}
\end{center}
\caption{Average value of $E^2(x,t)$ in arbitrary units in the
neighbourhood of $x=0.5$ at time $t=0.25$, with $10^4$ modes
(dashed line), $2 \cdot 10^4$ (dotted line) and $3 \cdot 10^4$
(continuous line). The distribution of the field modes used here
is not symmetric respect to the atomic transition frequency.}
\label{fig:3}
\end{figure}

The behaviour is quite different when the number of field modes is
increased in such a way that only the upper cut-off frequency
increases but the lower cut-off frequency remains fixed (in this
case the frequencies of the field modes are not symmetric with
respect to the atomic transition frequency); this situation is
intended to mimic the case in which the field modes extend from
$0$ to $\infty$, and no extension to $-\infty$ is performed. Fig.
\ref{fig:3} shows a zoom around $x=0.5$ of the square of the
electric field at time $t=0.25$ for $10^4$,  $2 \cdot 10^4$ and $3
\cdot 10^4$ modes. The figures indicate that, as the number of
modes is increased, the envelop of the oscillations does not
approximate a sharp front. The remaining tail gives a noncausal
behaviour. This makes quite evident that the use of the rotating
wave approximation with an Hamiltonian bounded from below does not
give causality in the propagation of the electromagnetic fields,
because noncausal tails persist, consistently with Hegerfeldt's
theorem.

\begin{figure}[ht]
\begin{center}
\includegraphics*[width=15cm]{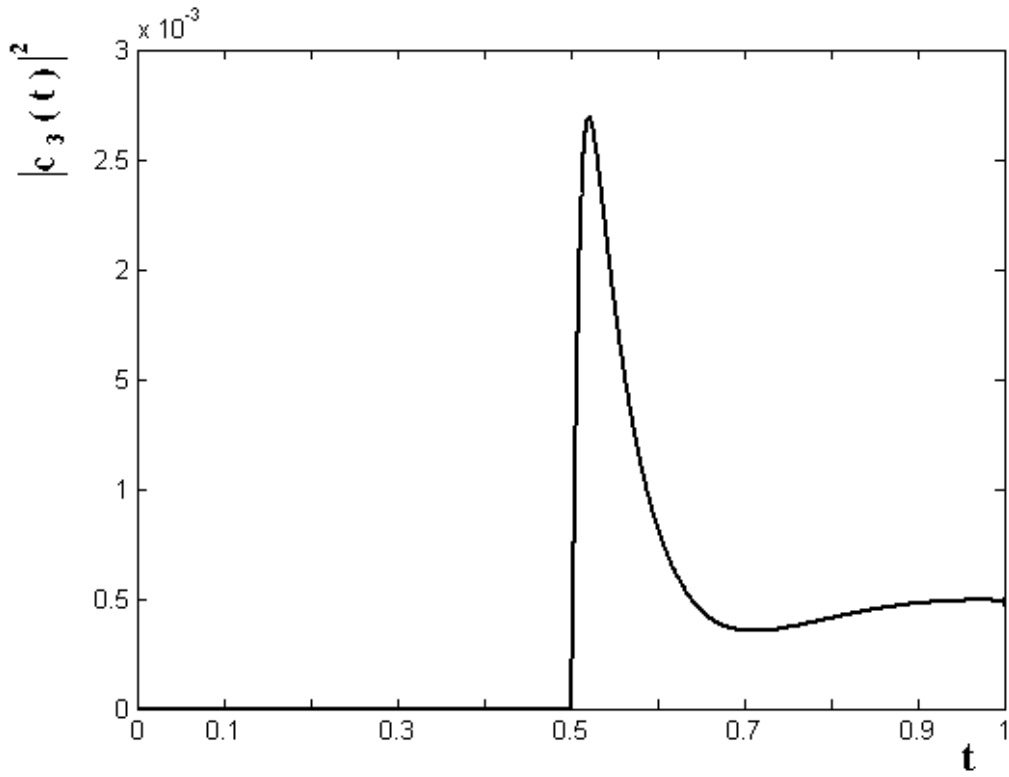}
\end{center}
\caption{Excitation probability of atom 3, with a configuration of
the field modes symmetric around $\omega_1$.} \label{fig:4}
\end{figure}

\begin{figure}[ht]
\begin{center}
\includegraphics*[width=15cm]{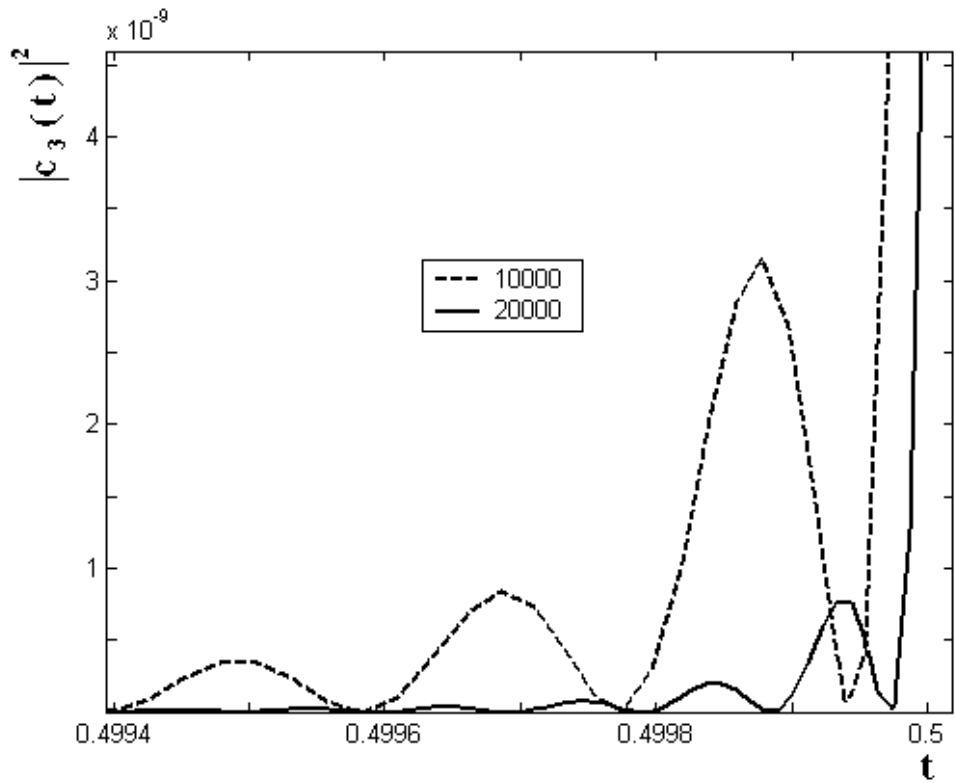}
\end{center}
\caption{Zoom of Fig. 4: excitation probability of atom 3 with a
symmetric configuration of the field modes, around the causality
time $t=0.5$.} \label{fig:5}
\end{figure}

\begin{figure}[ht]
\begin{center}
\includegraphics*[width=15cm]{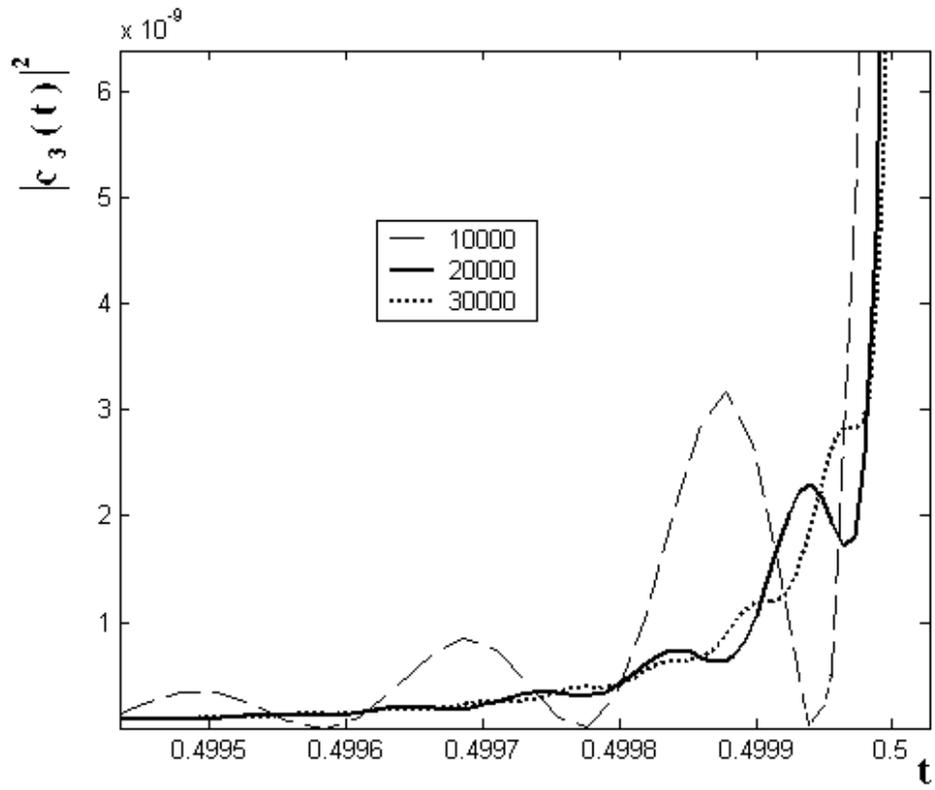}
\end{center}
\caption{Zoom of the excitation probability of atom 3 around the
causality time $t=0.5$, using a non-symmetric distribution of the
field modes.} \label{fig:6}
\end{figure}

Similar conclusions are reached by calculating $\mid c_3(t)
\mid^2$, that is the excitation probability of atom 3. Causality
requires that this probability should vanish for $t<0.5$. Fig.
\ref{fig:4} shows our numerical results, and Fig. \ref{fig:5} a
zoom around the {\it causality time} $t=0.5$ (with $10^4$ and $2
\cdot 10^4$ modes), both with a symmetric configuration of the
mode frequencies. We note that a sharper front is obtained when
the number of the modes is increased. However, Fig. \ref{fig:6}
shows the result for the non-symmetric configuration, in which
only the upper cut-off frequency is increased with the field
modes. It is evident from Fig. \ref{fig:6} that in this case the
noncausal tails for $t<0.5$ do not decrease on the average as the
number of the modes is increased.

We wish to conclude by stressing that our results clearly show the
reason why recent results in the literature have obtained a causal
behaviour of the atom-field interaction in a cavity within the
rotating wave approximation. The point is that at the same time
the frequency integration over the field modes was extended to
$-\infty$ which permits to escape the conditions set by
Hegerfeldt's theorem by making the Hamiltonian unbounded from
below, but which is physically unacceptable. In this paper, we
have considered three two-level atoms, one excited and two in the
ground state, inside a one-dimensional cavity, interacting with
the electromagnetic radiation field in the RWA. We have calculated
numerically the energy density of the electric field spontaneously
emitted by the excited atom and scattered by the second atom, as
well as the probability of excitation of the second and third
atom. We have shown that, without the (arbitrary) extension of the
field frequencies to $-\infty$ (frequently used in the
literature), and which has no physical basis, noncausal tails are
present both in the field propagation inside the cavity and in the
atomic excitation probabilities, even when the number of the modes
is increased. This underlines the potential dangers of the
rotating wave approximation. In a forthcoming paper, we will
explicitly show that the correct inclusion of the counter-rotating
terms of the Hamiltonian, allows to obtain a better causal
behaviour without any need of extending the frequency of the modes
to $-\infty$.

The authors wish to thank P.P. Corso for helpful comments and
suggestions on the numerical calculations. This work was in part
supported by the bilateral Italian-Belgian project on
``Casimir-Polder forces, Casimir effect and their fluctuations"
and the bilateral Italian-Japanese project 15C1 on ``Quantum
Information and Computation" of the Italian Ministry for Foreign
Affairs. Partial support by Ministero dell'Universit\`{a} e della
Ricerca Scientifica e Tecnologica and by Comitato Regionale di
Ricerche Nucleari e di Struttura della Materia is also
acknowledged.

\end{document}